\documentclass[aps,prd,reprint,preprintnumbers,showpacs,nofootinbib]{revtex4-1}

\usepackage[rgb]{xcolor}
\usepackage{color}
\usepackage{amssymb,amsmath,mathrsfs,graphicx,stmaryrd,bbm,dsfont,booktabs,mathtools,braket,lmodern,tikz}

\DeclareMathAlphabet{\mathpzc}{OT1}{pzc}{m}{it}

\DeclareMathAlphabet{\mathsc}{T1}{lmr}{m}{scsl}

\newcommand{\bea}{\begin{eqnarray}}
\newcommand{\eea}{\end{eqnarray}}
\def\be{\begin{equation}}
\def\ee{\end{equation}}
\newcommand{\bei}{\begin{itemize}}
\newcommand{\eei}{\end{itemize}}
\newcommand{\bee}{\begin{enumerate}}
\newcommand{\eee}{\end{enumerate}}

\def\z{\zeta}

\def\ads{{\rm AdS}_5\times {\rm S}^5}

\def\ads{{\rm AdS}_5\times {\rm S}^5}

\def\am{{\rm am}}
\def\am0{{\rm am}_0}

\DeclareMathOperator{\arcsinh}{arcsinh}

\expandafter\def\expandafter\bfseries\expandafter{\bfseries\ifmmode\else\boldmath\fi}
\expandafter\def\expandafter\mdseries\expandafter{\mdseries\ifmmode\else\unboldmath\fi}
\expandafter\def\expandafter\normalfont\expandafter{\normalfont\ifmmode\else\unboldmath\fi}

\definecolor{grey}{rgb}{0.4,0.4,0.5}
\definecolor{darkgreen}{rgb}{0,0.5,0}
\definecolor{darkred}{rgb}{0.6,0.0,0}
\definecolor{lightbrown}{rgb}{1,0.9,0.8}
\definecolor{brown}{rgb}{0.6,0.3,0.3}
\definecolor{darkblue}{rgb}{0,0,0.8}
\definecolor{darkmagenta}{rgb}{0.5,0,0.5}

\begin{document}

\title{Quantum deformations of the flat space superstring}
\author{Anna Pacho\l}
\email{apachol@unito.it}
\affiliation{Dipartimento di Matematica "Giuseppe Peano", Universit\`a degli
Studi di Torino, Via Carlo Alberto, 10 - 10123 Torino, Italy}
\author{Stijn J. van Tongeren}
\email{svantongeren@physik.hu-berlin.de}
\affiliation{Institut f\"ur Mathematik und Institut f\"ur Physik, Humboldt-Universit\"at zu Berlin, IRIS Geb\"aude, Zum Grossen Windkanal 6, 12489 Berlin, Germany}

\begin{abstract}
We discuss a quantum deformation of the Green-Schwarz superstring on flat space, arising as a contraction limit of the corresponding deformation of $\ads$. This contraction limit turns out to be equivalent to a previously studied limit that yields the so-called mirror model - the model obtained from the light cone gauge fixed $\ads$ string by a double Wick rotation. Reversing this logic, the $\ads$ superstring is the double Wick rotation of a quantum deformation of the flat space superstring. This quantum deformed flat space string realizes symmetries of timelike $\kappa$-Poincar\'e type, and is T dual to $\mathrm{dS}_5 \times \mathrm{H}^5$, indicating interesting relations between symmetry algebras under T duality. Our results directly extend to $\mathrm{AdS}_2 \times \mathrm{S}^2 \times \mathrm{T}^6$ and $\mathrm{AdS}_3 \times \mathrm{S}^3 \times \mathrm{T}^4$, and beyond string theory to many (semi)symmetric space coset sigma models, such as for example a deformation of the four-dimensional Minkowski sigma model with timelike $\kappa$-Poincar\'e symmetry. We also discuss possible null and spacelike deformations.
\end{abstract}


\preprint{HU-EP-15/48, HU-MATH-15/13}

\maketitle

Understanding the dynamics of a string in a generic background is a complicated problem. The simplest possible background for a string is flat space, and there its dynamics are well understood. In case of a background such as $\ads$ - playing an important role in the AdS/CFT correspondence \cite{Maldacena:1997re} - the sigma model is considerably more complicated, but can still be tackled thanks to its integrability \cite{Arutyunov:2009ga,Beisert:2010jr}. Given the computational control that integrability offers, efforts have been made to find other, less symmetric backgrounds that nevertheless correspond to an integrable model. One way to do so is to deform the string in a controlled fashion, an example of which is the Lunin-Maldacena deformation \cite{Lunin:2005jy,Frolov:2005dj,Frolov:2005ty}, the string dual to $\beta$ deformed supersymmetric Yang-Mills theory. A more drastic deformation was proposed in \cite{Delduc:2013qra}, corresponding to a quantum ($q$) deformation of the $\ads$ string sigma model in the sense that the superconformal algebra of the string is deformed to the corresponding quantum group \cite{Arutyunov:2013ega,Delduc:2014kha}.\footnote{This generalizes the earlier results of \cite{Klimcik:2002zj,Klimcik:2008eq,Delduc:2013fga}.} Based on experience with the squashed sphere sigma model \cite{Kawaguchi:2012ve} - which fits this framework \cite{Delduc:2013fga,Hoare:2014oua} - the full symmetry algebra of this model is expected to be the corresponding quantum affine algebra, a deformation of the Yangian of the $\ads$ string. The status of this deformed sigma model in terms of string theory remains mysterious to date \cite{Arutyunov:2015qva,Hoare:2015wia}\footnote{\label{footnote:defsugra}The fermions of the deformed model do not appear to be compatible with supergravity \cite{Arutyunov:2015qva}. At the same time, doing formal worldsheet T duality in all Cartan isometry directions does give a background compatible with supergravity \cite{Hoare:2015wia}. While this solution has a nontrivial dilaton that prevents T dualizing back in supergravity, this still means that the original deformed model should at least be scale invariant at one loop \cite{Hoare:2015wia}. Furthermore, at the level of scattering theory it seems desirable to do a nontrivial change of basis \cite{Arutyunov:2015qva}, which may have further consequences.} (see also \cite{Hoare:2014pna,Hoare:2015gda}). In this paper we investigate $q$ deformations in string theory in the simplest possible setting, that of flat space.

As we will explain, the deformed flat space string is intimately connected to the $\ads$ string in two ways. The first of these is in its construction. Semisimple Lie (super)algebras can be naturally deformed to quantum groups \cite{Chari}, but this procedure cannot be applied to nonsemisimple algebras such as the ten-dimensional (super) Poincar\'e algebra of the flat space string. However, given a suitable semisimple algebra it is possible to get a nontrivial deformation of some nonsemisimple algebras by the analogue of a Wigner-\.{I}n\"on\"u contraction. The result can be thought of as a quantum deformation of the corresponding nonsemisimple algebra. This was famously done for the four-dimensional Poincar\'e algebra, yielding what is known as the $\kappa$-Poincar\'e algebra from the $q$-deformed $\mathrm{AdS}_4$ algebra \cite{Lukierski:1991pn,Lukierski:1992dt}. Now, since the flat space string arises from a Wigner-\.{I}n\"on\"u type contraction of the $\ads$ one, it should be possible to obtain a $q$ deformation of the flat space string by an appropriate contraction of the $q$-deformed $\ads$ sigma model. The resulting symmetry algebra is of timelike (super) $\kappa$-Poincar\'e type. The implementation of this contraction yields the second connection to the $\ads$ string. It turns out that this contraction is nothing but a previously studied limit of the $q$-deformed $\ads$ string \cite{Arutynov:2014ota,Arutyunov:2014cra,Arutyunov:2014jfa}, related to the so-called mirror model \cite{Arutyunov:2007tc}. More precisely, this limit gives a sigma model that in a light cone gauge fixed setting is related to the $\ads$ one by a worldsheet double Wick rotation.\footnote{\label{footnote:mirrorferms}Cf. footnote \ref{footnote:defsugra}, explicitly matching the fermions of this mirror background \cite{Arutyunov:2014cra,Arutyunov:2014jfa} with those of the contracted deformed sigma model is a subtle point: a direct limit does not appear to give the desired answer, however after a suitable change of basis of scattering states the associated S matrices do exactly match \cite{Arutyunov:2015qva}.} In other words, the $\ads$ string is the double Wick rotation of the $q$ deformation of the simplest possible string.

In this paper we consider the bosonic sector of the model - where we go from $\mathcal{U}_q(\mathfrak{so}(2,4)\oplus \mathfrak{so}(6))$ to $\mathcal{U}_\kappa (\mathfrak{iso}(1,4) \oplus \mathfrak{iso}(5))$ symmetry - leaving a detailed investigation of fermions for the future. We do however match the well known ``lattice'' or ``spin chain'' $\sin P/2$ off shell central extension of $\mathfrak{psu}(2|2)$ \cite{Beisert:2005tm,Arutyunov:2006ak} that plays an important role in integrability in AdS/CFT. From the present point of view, it is the double Wick rotation of the contraction of a fermionic anticommutator in $\mathcal{U}_q(\mathfrak{psu}(2,2|4))$. We tentatively refer to this only implicitly described contraction of $\mathcal{U}_q(\mathfrak{psu}(2,2|4))$ as (inhomogeneous)  $\mathcal{U}_\kappa(\mathfrak{iusp}(2,2|4))$ (note that $\mathfrak{usp}(2,2) \simeq \mathfrak{so}(4,1)$ and $\mathfrak{usp}(4) \simeq \mathfrak{so}(5)$).

The metric of the $q$-deformed flat space string, also known as mirror $\ads$, is related to $\mathrm{dS}_5 \times \mathrm{H}^5$ by two T dualities, one involving time. Now, the type IIB$^*$ sigma model on $\mathrm{dS}_5 \times \mathrm{H}^5$ has $\mathfrak{su}^*(4|4)$ symmetry \cite{Hull:1998vg}, which should extend to a full Yangian algebra based on $\mathfrak{su}^*(4|4)$. Our model on the other hand has $\mathcal{U}_\kappa(\mathfrak{iusp}(2,2|4))$ symmetry, which we expect to extend to the corresponding quantum affine algebra. Double T duality hence appears to relate (two realizations of) these infinite-dimensional symmetry algebras.\footnote{\label{footnote:reftootherfootnotes}This assumes the subtleties with the fermions of the deformed model indicated in footnotes \ref{footnote:defsugra} and \ref{footnote:mirrorferms} can be appropriately resolved.} At the bosonic level, the timelike T duality apparently relates (the infinite-dimensional extensions of) $\mathfrak{so}(1,5)$ and $\mathcal{U}_\kappa(\mathfrak{iso}(1,4))$  while the spacelike one relates $\mathfrak{so}(1,5)$ and $\mathcal{U}_\kappa(\mathfrak{iso}(5))$.

Our construction and these comments readily generalize to other dimensions, in particular to superstrings on $\mathrm{AdS}_2 \times \mathrm{S}^2 \times \mathrm{T}^6$ and $\mathrm{AdS}_3 \times \mathrm{S}^3 \times \mathrm{T}^4$. These would present different $q$ deformations of the flat space string, with smaller isometry subgroups being deformed. In this sense then, our contraction of the deformed $\ads$ string gives the largest possible deformation in flat space. At the level of bosonic sigma models we can consider separate spaces and many dimensions, which makes it possible to realize null and spacelike $\kappa$-Poincar\'e type symmetry, and to for instance make contact with the four-dimensional $\kappa$-Poincar\'e algebra by considering an analogous contraction of the $q$-deformed $\mathrm{AdS}_4$ sigma model.

This paper is organized as follows. We will begin by briefly introducing contractions of quantum algebras. Then we implement this type of contraction in the sigma model - demonstrating that the standard Drinfeld-Jimbo $r$ matrix for $\mathcal{U}_q(\mathfrak{so}(2,4)\oplus \mathfrak{so}(6))$ reduces to the expected $\kappa$-Poincar\'e type $r$ matrix - and indicate its relation to the mirror model. Next we discuss the off shell central extension of $\mathfrak{psu}(2|2)$, and comment on T duality relations. We then comment on bosonic sigma models in various dimensions and the associated spacelike and null type contractions. In the conclusions we indicate possible generalizations and open questions. Appendices contain a discussion of the relevant Lie algebras and $r$ matrices, as well as comments on two other possible deformations of $\ads$, deformations of $\mathrm{dS}_5 \times \mathrm{H}^5$, and deformations of $\mathrm{S}^2$, $\mathrm{AdS}_2$, $\mathrm{dS}_2$, and $\mathrm{H}^2$.

\section{Quantum algebra contraction}

The sigma models that we are considering have quantum group symmetry. At the bosonic level the relevant undeformed algebras are $\mathfrak{so}(2,d-1)$, $\mathfrak{so}(1,d)$, and $\mathfrak{so}(d+1)$, symmetries of anti-de Sitter space, de Sitter space and the hyperboloid, and the sphere, all $d$-dimensional. The essence of the contraction we are interested in is already captured in the simple case of two dimensions, which we would briefly like to recall. A clear pedagogical discussion of this topic can be found in \cite{Ruegg:1993eq}.

The quantum algebras $\mathcal{U}_q(\mathfrak{so}(3)) \simeq \mathcal{U}_q(\mathfrak{su}(2)) $ and $\mathcal{U}_q(\mathfrak{so}_q(2,1)) \simeq \mathcal{U}_q(\mathfrak{su}(1,1))$ are two relevant real forms of $\mathcal{U}_q(\mathfrak{sl}(2,\mathbb{C}))$ that are naturally defined with $q$ taken real.\footnote{By naturally we mean we are dealing with a standard notion of conjugation \cite{Ruegg:1993eq}.} $\mathcal{U}_q(\mathfrak{sl}(2,\mathbb{C}))$ is given by\footnote{We consider only the algebraic sector of the full Hopf algebra in the present paper.}
\begin{equation}
[e_+,e_-]=[h]_q, \,\,\, [h,e_\pm]=\pm2e_\pm,
\end{equation}
where
\begin{equation}
[a]_q = \frac{q^a - q^{-a}}{q - q^{-1}}.
\end{equation}
In terms of the physical antihermitian generators we use below, for $\mathcal{U}_q(\mathfrak{su}(2))$ we have
\begin{equation}
\begin{aligned}
[n_{16},n_{15}]&=\frac{i}{2}[-2i n_{56}]_q,\\
[n_{15},n_{56}]&=n_{16},\\
[n_{56},n_{16}]&=n_{15},
\end{aligned}
\end{equation}
while for $\mathcal{U}_q(\mathfrak{su}(1,1))$ instead
\begin{equation}
\begin{aligned}
[m_{15},m_{10}]&=\frac{i}{2}[-2i m_{05}]_q,\\
[m_{10},m_{05}]&=-m_{15},\\
[m_{05},m_{15}]&=-m_{10}.
\end{aligned}
\end{equation}
These are related by the analytic continuation $n_{16}=i m_{10}$, $n_{15}=i m_{15}$, $n_{56}=m_{05}$.\footnote{\label{footnote:Uqsl2}While isomorphic at the undeformed level, as nicely explained in \cite{Ruegg:1993eq}, there is a third natural deformation, $\mathcal{U}_q (\mathfrak{sl}(2))$ for $|q|=1$. The difference with $\mathcal{U}_q (\mathfrak{su}(1,1))$ is that in this case one deforms the commutator with a noncompact direction on the right hand side.} The conventional Wigner-\.{I}n\"on\"u contraction of $\mathfrak{so}(3)$ to the two-dimensional Euclidean algebra starts with the splitting $\mathfrak{so}(3) = \mathfrak{so}(2) \oplus \mathfrak{n}_2$ as vector spaces. We choose this $\mathfrak{so}(2)$ to be generated by $n_{15}$, so that the complement $\mathfrak{n}_2$ is spanned by $n_{16}$ and $n_{15}$. For the contraction to the Poincar\'e algebra we split $\mathfrak{so}(2,1) = \mathfrak{so}(1,1) \oplus \mathfrak{m}_2$, with the $\mathfrak{so}(1,1)$ generated by $m_{01}$, and hence the complement $\mathfrak{m}_2$ spanned by $m_{05}$ and $m_{15}$. We then rescale the generators in  $\mathfrak{n_2}$ and $\mathfrak{m_2}$ as $n_{i6}=R l_i,i=1,5$, and $m_{j5}=R p_j,j=0,1$. To keep a nontrivial deformation we also scale $q$ as $\log q = -\alpha/R$  \cite{Celeghini:1990xx}. In the limit $R\rightarrow \infty$ we then get
\begin{equation}
\label{eq:so3contr}
\begin{aligned}
[l_1,n_{15}]&=\frac{\sin 2 \alpha l_{5}}{2\alpha},\\
[n_{15},l_5]&=l_{1},\\
[l_{5},l_1]&=0,
\end{aligned}
\end{equation}
and
\begin{equation}
\label{eq:so12contr}
\begin{aligned}
[p_{1},m_{10}]&=\frac{\sin 2 \alpha p_{0}}{2\alpha},\\
[m_{10},p_{0}]&=-p_{1},\\
[p_{0},p_{1}]&=0,
\end{aligned}
\end{equation}
respectively. In the limit $\alpha\rightarrow 0$ these clearly reduce to the two-dimensional Euclidean and Poincar\'e algebras $\mathfrak{iso}(2)=\mathfrak{so}(2)\niplus \mathfrak{l}_2$ and $\mathfrak{iso}(1,1)=\mathfrak{so}(1,1)\niplus \mathfrak{p}_2$, with $\mathfrak{l}_2$ and $\mathfrak{p}_2$ generated by the $l$s and $p$s. For $\mathfrak{so}(2,1)$ we could have also chosen the splitting $\mathfrak{so}(2) \oplus \mathfrak{m}^\prime_2$, but in that case we would trivialize the deformation.

Though more involved, higher-dimensional algebras can be similarly contracted by appropriately splitting (real forms of) $\mathfrak{so}(n+1)$ into an $\mathfrak{so}(n)$ factor and an appropriate complement $\mathfrak{m}_n$, and scaling the $n$ generators in $\mathfrak{m}_n$ together with an appropriate scaling of $q$ at the level of the corresponding quantum algebra. Contracting $\mathcal{U}_q (\mathfrak{so}(2,3))$ this way gives the famous $\kappa$-Poincar\'e algebra \cite{Lukierski:1991pn,Lukierski:1992dt} for example. Higher rank special orthogonal algebras have not been explicitly contracted, but all should yield $\kappa$-Poincar\'e type deformations of the associated flat space isometry groups of appropriate signature, which do exist in higher dimensions \cite{Lukierski:1993wxa,Borowiec:2013lca}. As we will come back to below, the so-called $r$ matrices associated with this type of deformations are of the form
\begin{equation}
\label{eq:rkappageneral}
r = a^\mu \hat{m}_{\mu\nu} \wedge p^\nu,
\end{equation}
where the $\hat{m}_{\mu\nu}$ generate an appropriate real form of $\mathfrak{so}(n)$, $p_\nu$ are the translation generators, and in case of indefinite signature $a^\mu$ can be a timelike, spacelike, or null vector. The contraction we described for $\mathcal{U}_q(\mathfrak{so}(2,1))$ results in the two-dimensional analogue of the timelike $\kappa$-Poincar\'e algebra.

The idea is now to implement this type of contraction in the $q$-deformed $\ads$ ($\mathrm{AdS}_n \times \mathrm{S}^n \times \mathrm{T}^{10-2n}$) string sigma model, which will (hopefully) give a nontrivial deformation of the flat space string with $\kappa$-Poincar\'e type symmetry.

\section{Contracting the sigma model}

In this section we will discuss the implementation of the contraction procedure described above in the sigma model. We will focus on the bosonic sector of the model.

\subsection{The sigma model}

The action for the bosonic string is given by
\begin{equation}
\label{eq:stringaction}
S=-\tfrac{T}{2}\int{\rm d}\tau{\rm d}\sigma \, \left(g_{\mathsc{mn}}\, dx^\mathsc{m} dx^\mathsc{n} - B_{\mathsc{mn}}\, dx^\mathsc{m} \hspace{-3pt} \wedge dx^\mathsc{n}\right) ,
\end{equation}
where $T$ is the string tension, and $g$ and $B$ denote the background metric and $B$ field respectively. The string action for $\ads = \mathrm{SO}(2,4)/\mathrm{SO}(1,4) \times \mathrm{SO}(6)/\mathrm{SO}(5)$ can be written as a (semi)symmetric space coset sigma model \cite{Metsaev:1998it}, which can be deformed based on a so-called $\mathrm{R}$ operator as proposed in \cite{Delduc:2013qra}, giving a so-called Yang-Baxter sigma model \cite{Klimcik:2002zj,Klimcik:2008eq}. The bosonic action is\footnote{Here $h$ is the world sheet metric, $\epsilon^{\tau\sigma}=1$, $A_\alpha = g^{-1} \partial_\alpha g$ with $g\in \mathrm{PSU}(2,2|4)$, $\mathrm{sTr}$ denotes the supertrace, and the $P_i$ are the projectors onto the $i$th $\mathbb{Z}_4$ graded components of the semi-symmetric space $\mathrm{PSU}(2,2|4)/(\mathrm{SO}(4,1)\times \mathrm{SO}(5))$ (super $\ads$).}
\begin{equation}
\label{eq:defaction}
S = -\tfrac{T}{2} \int d\tau d\sigma \tfrac{1}{2}(\sqrt{h} h^{\alpha \beta} -\epsilon^{\alpha \beta}) \mathrm{sTr} (A_\alpha P_2 J_\beta)
\end{equation}
where $J=(1-\varkappa \mathrm{R}_g \circ P_2)^{-1}(A)$ with $\mathrm{R}_g(X)=g^{-1} \mathrm{R}(g Xg^{-1}) g$, and $\varkappa$ labels the deformation. At $\varkappa=0$ this gives the undeformed $\ads$ string. The operator $\mathrm{R}$ is a skew symmetric map from a relevant (super)algebra to itself, here $\mathfrak{su}(2,2|4)$, which solves the modified classical Yang-Baxter equation (mCYBE)
\begin{equation}\label{eq:ROpmCYBE}
[\mathrm{R}(x),\mathrm{R}(y)]-\mathrm{R}([\mathrm{R}(x),y]+[x,\mathrm{R}(y)])=\pm[x,y].
\end{equation}
In this case the $\mathrm{R}$ operator is of the so-called nonsplit type, meaning it solves the mCYBE with a $+$ sign. For completeness, a $-$ sign means a split solution, and dropping the commutator on the right hand side altogether gives the homogeneous classical Yang-Baxter equation (CYBE).\footnote{See \cite{Vicedo:2015pna} for a recent unified discussion of these three classes of deformations in the present context.}

Including fermions, this deformed model realizes $\mathcal{U}_q(\mathfrak{psu}(2,2|4))$ symmetry \cite{Delduc:2014kha}, where the deformation parameter $q$ is (classically) expressed in terms of the string tension $T$, the $\mathrm{AdS}_5$ radius of curvature $R$, and the deformation parameter $\varkappa$ as \cite{Arutyunov:2013ega}
\begin{equation}
\label{eq:q}
\log q = -\varkappa/(R^2 T).
\end{equation}

The effect of this deformation on the bosonic background was worked out in \cite{Arutyunov:2013ega} (for fermions see \cite{Arutyunov:2015qva}), and the result is\footnote{We hope the distinction between the operator $\mathrm{R}$ and the physical scale $R$ is clear.}
\begin{align}
\nonumber
R^{-2} ds^2 = \,&
-\frac{f_+(\rho)}{f_-(\varkappa \rho)} dt^2 +\frac{1}{f_+(\rho)f_-(\varkappa \rho)} d\rho^2 + \rho^2 d\Theta^\rho_{3}\\
&
 + \frac{f_-(r)}{f_+(\varkappa r)} d\phi^2 +\frac{1}{f_-(r)f_+(\varkappa r)} dr^2 +r^2 d\Theta^r_{3},\nonumber
\end{align}
where $f_{\pm}(x) = 1\pm x^2$, $d\Theta_{3}$ is a deformation of the three-sphere metric in Hopf coordinates
\begin{align}
d\Theta^\rho_{3} \equiv & \, \frac{1}{1+\varkappa^2 \rho^4 \sin^2 \zeta} (d\zeta^2 + \cos^2 \zeta d\psi_2^2) +  \sin^2 \zeta d \psi_3^2,\nonumber
\\
d\Theta^r_{3} \equiv & \, \frac{1}{1+\varkappa^2 r^4 \sin^2 \xi} (d\xi^2 + \cos^2 \xi d\chi_2^2) +  \sin^2 \xi d \chi_3^2.\nonumber
\end{align}
At $\varkappa=0$ this is the metric of $\ads$, whose factors have curvature $-20/R^2$ and $20/R^2$ respectively. The $B$ field is given by
\begin{align}
R^{-2} B = &\, \varkappa \Big(\frac{\rho ^4 \sin 2 \zeta}{1+ \varkappa ^2 \rho ^4 \sin ^2\z} d\psi_1 \wedge d\z \nonumber\\
& \hspace{1.2cm} -\frac{ r^4 \sin 2 \xi }{1+ \varkappa ^2 r^4 \sin^2\xi} d\chi_1 \wedge d\xi\Big).\nonumber
\end{align}
These expressions arise by deforming the $\ads$ coset sigma model built on $g= \mbox{diag}(g_a,g_s)$ with
\begin{equation}
\label{eq:groupelement}
\begin{aligned}
 g_a & = e^{\psi_i f^i} e^{- \zeta m^{13}} e^{ \arcsinh{\rho}\, m^{15}},\\
 g_s & = e^{\chi_i f^i} e^{- \xi n^{13}} e^{\arcsin{r}\, n^{16}},
\end{aligned}
\end{equation}
where $t=\psi_1$ and $\phi=\chi_1$, and the $f^i$ are given by $f^1 = m^{05}=n^{56}$, $f^2 = m^{12}=n^{12}$ and $f^3 = m^{34}=n^{34}$, which span the Cartan subalgebra of $\mathfrak{psu}(2,2|4)$. Our conventions for $\mathfrak{so}(2,4)$ and $\mathfrak{so}(6)$ and their generators $m^{ij}$ and $n^{ij}$ are discussed in appendix \ref{app:algebraandrmatrix}. As will come back later, there are multiple choices of $\mathrm{R}$ operator. The above background corresponds to the standard $\mathrm{R}$ operator, associated with the standard Drinfeld-Jimbo $r$ matrix as discussed in appendix \ref{app:algebraandrmatrix}.

\subsection{The contraction}

Now we are ready to consider contractions. Analogously to $\mathfrak{so}(2,1)$ discussed above where we split off $\mathfrak{so}(1,1)$, for $\mathfrak{so}(2,4)$ we want to split off an $\mathfrak{so}(1,4)$ algebra. Importantly, we take this $\mathfrak{so}(1,4)$ to be the algebra of the coset denominator, which is generated by $m$s with indices running from zero through four, so that at the undeformed level we contract $\mathrm{SO}(2,4)/\mathrm{SO}(1,4)$ to $\mathrm{ISO}(1,4)/\mathrm{SO}(1,4) \simeq \mathbb{R}^{1,4}$. For $\mathfrak{so}(6)$ we follow the same procedure. In other words, in the split relevant for the contraction, the complement to $\mathfrak{so}(1,4)$ in $\mathfrak{so}(2,4)$ is spanned by the $m^{a5}$, while the complement to $\mathfrak{so}(5)$ in $\mathfrak{so}(6)$ is spanned by $n^{c6}$.

Via the group parametrization of eqs. \eqref{eq:groupelement} we can translate the contraction we are interested in to the coordinates. This amounts to a singular limit on $t$ and $\rho$, and $r$ and $\phi$.\footnote{Had we not chosen our splitting compatible with the coset structure, we would formally need to rescale directions corresponding to gauge degrees of freedom. It is not clear to us whether this can be sensibly done.} We can physically implement this by reinstating dimensionful fields through the $\mathrm{AdS}_5$ radius as
\begin{align}
\nonumber
\tilde{t} & = R t,  & \tilde{\phi} & = R \phi, & \tilde{r} & = R \rho,  & \tilde{\rho} & = R r,
\end{align}
and taking the limit $R\rightarrow \infty$ keeping the new coordinates fixed. Here we relabeled $\rho$ and $r$ to smoothly match established results below. If we do not do anything else, this limit gives flat space, matching the algebraic situation discussed above, where to keep a nontrivial deformation we had to simultaneously scale $q$. Given eqn. \eqref{eq:q}, it is actually natural to do so. We can consider the limit $R,\varkappa\rightarrow \infty$, keeping $\varkappa/R \equiv \kappa^{-1}$ and the string tension fixed.\footnote{Note the distinction between $\varkappa$ and $\kappa$ here. These two variables are standard in their respective fields (deformed string sigma models and $\kappa$-Poincar\'e algebras), so we felt it better not to introduce further ones.} In this limit $q\rightarrow1$, while the symmetry algebra should remain partially deformed as is clear from the simple examples discussed in the previous section. In this limit the $B$ field vanishes, while the metric becomes
\begin{align}
ds^2 = \,&
\frac{-dt^2  + d r^2}{1-r^2/\kappa^{2}} + r^2 d\Theta_{3}\label{eq:mirrorwithkappa}\\
&
 +\frac{d\phi^2  + d \rho^2}{1+\rho^2/\kappa^{2}} +\rho^2 d\Theta_{3},\nonumber
\end{align}
where we have dropped tildes - note that the range of $\phi$ is no longer compact. Similar to the radius of $\mathrm{AdS}_5$, $\kappa$ simply sets an overall scale, and factoring it out by rescaling coordinates, we get a string on
\begin{align}
ds^2 = \,&
\frac{-dt^2  + d r^2}{1-r^2} + r^2 d\Theta_{3}\label{eq:mirrorwithoutkappa}\\
&
 +\frac{d\phi^2  + d \rho^2}{1+\rho^2} +\rho^2 d\Theta_{3},\nonumber
\end{align}
with effective string tension $g \equiv \kappa^{2} T$.

At this stage the model may appear to contain no deformation parameter anymore, and indeed the deformation parameter can be formally scaled out of $\kappa$-Poincar\'e type algebras, like it can be from eqs. \eqref{eq:so3contr} and \eqref{eq:so12contr}.\footnote{This is possible in the so-called $q$-analog version of $\kappa$-Poincar\'e type algebras, where the value of $\kappa$ can be fixed, see e.g. \cite{Borowiec:2010yw}.} However, this is entirely analogous to how the $\ads$ radius can be scaled out of $\mathfrak{psu}(2,2|4)$ and absorbed in the string tension. The value of the deformation parameter becomes relevant as soon as a physical scale is fixed.\footnote{At the level of the quantum spectrum the distinction is also clear: the flat space string has a fixed integer spectrum, up to an overall scale set by the string tension. Our string has a complicated spectrum depending on a dimensionless parameter $g$, with an overall scale set by either $\kappa$ or $T$. This is of course one fewer free parameter than the $q$-deformed $\ads$ model.}

As indicated in the introduction, this contraction should result in a symmetry algebra of the $\kappa$-Poincar\'e type. Such contractions have not been explicitly worked for our rank three cases $\mathcal{U}_q (\mathfrak{so}(2,4))$ and $\mathcal{U}_q (\mathfrak{so}(6))$, let alone $\mathcal{U}_q (\mathfrak{psu}(2,2|4))$. In all lower-dimensional cases ($\mathrm{AdS}_3 \times \mathrm{S}^3 \times \mathrm{T}^{4}$, $\mathrm{AdS}_2 \times \mathrm{S}^2 \times \mathrm{T}^{6}$, and e.g. the sigma model on $\mathrm{AdS}_4$) however, these contractions are exactly the ones that give $\kappa$-Poincar\'e type algebras \cite{Lukierski:1992dt}, leaving little doubt what the outcome should be. Still, to show this more concretely, rather than contracting the algebras in painstaking detail, let us focus on the associated $r$ matrices. As explained in detail in appendix \ref{app:algebraandrmatrix}, if we take the canonical Drinfeld-Jimbo $r$ matrix for the associated quantum groups, express it in terms of physical generators, and take the contraction limit appropriate for the sigma model, we get
\begin{equation}
r =  m_{0j} \wedge p^j , \mbox{ and } r =  n_{5j} \wedge l^j,
\end{equation}
for $\mathfrak{so}(2,4)$ and $\mathfrak{so}(6)$ respectively, sums in $j$ running from one through four. Cf. eqn. \eqref{eq:rkappageneral}, these are precisely the $\kappa$-Poincar\'e type $r$ matrices for $\mathcal{U}_\kappa(\mathfrak{iso}(1,4))$ and $\mathcal{U}_\kappa(\mathfrak{iso}(5))$ respectively \cite{Zakrzewski:1994,Borowiec:2013lca}, timelike in the case of $\mathfrak{iso}(1,4)$.\footnote{The indices $0$ on $m$ and $5$ on $n$ are entirely due to conventions, though it is relevant that the index on $m$ is timelike.} In other words, also these cases correspond to standard $\kappa$-Poincar\'e type algebras that can be found in \cite{Lukierski:1993wxa}. To extend these results to the supersymmetric case, we should include the supercharges and rescale them by $\sqrt{R}$ in the contraction procedure. We will not discuss this in detail in the present paper, but will come back to one nice aspect of it below.

Our deformed flat space string thus has an internal symmetry algebra of the $\kappa$-Poincar\'e type. This algebra acts on the string fields defined on the worldsheet, where its generators are realized via conserved charges. The metric and other background fields determine the worldsheet field theory interaction terms - hence symmetries - and it is precisely these that get deformed. This is different from the typical realization of the $\kappa$-Poincar\'e algebra as a noncommutative spacetime symmetry algebra, where the $\kappa$-Poincar\'e algebra is realized in terms of differential operators action on a noncommutative version of Minkowski space \cite{Majid:1994cy} called $\kappa$-Minkowski space. These are simply two different modules for the same algebra.\footnote{The AdS/CFT correspondence can ``mix'' these concepts, as e.g. the global conformal spacetime symmetry of $\mathcal{N}=4$ SYM corresponds to the global internal symmetry of the $\mathrm{AdS}_5$ sigma model. As such, we might expect deformed internal symmetries of the string to relate to deformed spacetime symmetries in the dual field theory. This expectation has been made more precise for so-called non-standard quantum deformations (Drinfeld twists) of $\ads$, which can include four-dimensional $\kappa$-Minkowski type structures in the dual field theory \cite{vanTongeren:2015uha}.}

\subsection{$q$ deformation as the mirror model}

The space of eqn. \eqref{eq:mirrorwithoutkappa} and the limit to get there were already considered from a different angle in \cite{Arutyunov:2014cra,Arutyunov:2014jfa}, where this background including a corresponding dilaton and Ramond-Ramond five form was shown to correspond to the so-called $\ads$ mirror model - a double Wick rotation of the light cone gauge fixed $\ads$ sigma model \cite{Arutyunov:2007tc}. For completeness, this dilaton and five form are \cite{Arutyunov:2014cra}
\begin{equation}\nonumber
\Phi=\Phi_0-\frac{1}{2}\log  (1-r^2)(1+\rho^2),
\end{equation}
and
\begin{equation}\nonumber
 F =  4 e^{-\Phi} \left( \omega_\phi - \omega_t \right).
\end{equation}
where $\Phi_0$ is a constant, and $\omega_t$ and $\omega_\phi$ denote would-be volume forms on the two five-dimensional submanifolds, except with $t$ and $\phi$ formally interchanged. While the light cone gauge fixed lagrangians of these two theories are related by a double Wick rotation, note that in particular the Virasoro constraints are different due to the interchange of space and time.

By this relationship between the contraction limit and the mirror model, the $\ads$ string is the (off shell) double Wick rotation of the $q$ deformation of the flat space superstring.\footnote{Admittedly, the $q$ deformation itself is found through the $\ads$ model to begin with.} It also shows that the symmetry algebra of the mirror $\ads$ string is $\mathcal{U}_\kappa (\mathfrak{iso}(1,4) \oplus \mathfrak{iso}(5))$ at the bosonic level. Modulo the subtleties mentioned earlier, upon including fermions this should extend to what we might denote as $\mathcal{U}_\kappa(\mathfrak{iusp}(2,2|4))$, though we have not explicitly described this deformed superalgebra here. Let us however discuss one aspect of this deformed superalgebra, where we can nicely make contact with well known aspects of the off-shell symmetry algebra of the $\ads$ string.

\subsection{Mirror fermions and off shell central extensions}

As double Wick rotations preserve conservation laws, the mirror string inherits the light cone symmetries of the $\ads$ superstring. The on shell symmetry algebra of the light cone gauge fixed $\ads$ string is centrally extended $\mathfrak{psu}(2|2)^{\oplus2}$, where the central element $\mathbb{H}$ corresponds to the worldsheet Hamiltonian. Focusing on one copy of $\mathfrak{psu}(2|2)$, the supercharges $Q$ and $Q^\dagger$ satisfy
\begin{align}
\nonumber
\{Q_{\alpha}^{\,\,a}, Q^{\dagger \beta}_b \} = \delta^a_b R_\alpha^{\,\,\beta} + \delta_\alpha^\beta L_b^{\,\,a} + \frac{1}{2} \delta^a_b \delta_\alpha^\beta \mathbb{H},
\end{align}
where $L$ and $R$ generate the two bosonic $\mathfrak{su}(2)$s. Note that this is a conventional Lie superalgebra. If we go off shell by relaxing the Virasoro constraint (level matching condition), this algebra picks up a further central element of the form $g \sin P/2$, where $P$ denotes the total worldsheet momentum \cite{Arutyunov:2006ak}, and $g$ is the effective $\ads$ string tension. Doing a double Wick rotation interchanges energy and momentum, and for the mirror theory we instead have \cite{Arutyunov:2007tc}
\begin{align}
\label{eq:mirrorpsu22}
\{\tilde{Q}_{\alpha}^{\,\,a}, \tilde{Q}^{\dagger \beta}_b \} = \delta^a_b R_\alpha^{\,\,\beta} + \delta_\alpha^\beta L_b^{\,\,a} + g \delta^a_b \delta_\alpha^\beta \sinh{\frac{\tilde{\mathbb{H}}}{2}},
\end{align}
where tildes denote mirror quantities, and we have put the mirror theory on shell by setting $\tilde{P}$ to zero. Here we see a signature of a would-be $q$ deformation: a hyperbolic sine.

To match the above with the contraction of $\mathcal{U}_q(\mathfrak{psu}(2,2|4))$, let us consider the simpler $\mathcal{U}_q(\mathfrak{psu}(1,1|2))$ instead, either as a subalgebra (before deformation), or as the relevant superalgebra for the $\mathrm{AdS}_2$ string where the same central extension appears, see e.g. \cite{Hoare:2015kla}. Standard light cone gauge fixing here produces two $\mathfrak{psu}(1|1)$ subalgebras with appropriate central extensions that couple them. These are given by the centralizer in $\mathfrak{psu}(2|2)$ with respect to $\mbox{diag}(1,-1,1,-1)$ - in other words one $\mathfrak{psu}(1|1)$  involves rows and columns one and three, the other two and four. Focusing on the first of these $\mathfrak{psu}(1|1)$s, its supercharges simply anticommute to $\bar{h}_1 + \bar{h}_3 = \mbox{diag}(1,0,1,0)$, which upon $q$ deformation becomes
\begin{equation}
\{Q,Q^\dagger\} = [\bar{h}_1+\bar{h}_3]_q,
\end{equation}
where $Q$ and $Q^\dagger$ are the supercharges of $\mathfrak{psu}(1|1)$. Since we are really dealing with $\mathcal{U}_q(\mathfrak{psu}(1,1|2)))$, at this stage we should set the overall central element of $\mathcal{U}_q(\mathfrak{su}(1,1|2)))$ to zero. Doing so means $2(\bar{h}_1+\bar{h}_3) \sim \mbox{diag}(1,-1,1,-1) = \mathbb{H}$ - the light cone string Hamiltonian. To do the quantum contraction we should rescale this generator by $R$ (it is the analogue of $i(m^{05}-n^{56})$), and as mentioned earlier the fermions by $\sqrt{R}$, which gives
\begin{equation}
\{Q,Q^\dagger\} = \kappa T \sinh{\frac{\mathbb{H}}{2\kappa T}}.
\end{equation}
This matches perfectly with the purely diagonal term of eqs. \eqref{eq:mirrorpsu22}, upon noting firstly that to get the mirror background in the standard form of eqn. \eqref{eq:mirrorwithoutkappa} - which eqs. \eqref{eq:mirrorpsu22} refer to -  we have to rescale $t\rightarrow\tilde{t}/\kappa$, meaning $\mathbb{H} \rightarrow \tilde{\mathbb{H}}= \kappa \mathbb{H}$, and analogously $Q\rightarrow \tilde{Q}= \sqrt{\kappa}Q$, and secondly that the conventions under which the off shell central extension was computed involve normalizing spatial translations by the string tension - see e.g. section 2.2.3 of \cite{Arutyunov:2009ga} - which we implicitly do not do here since everything remains associated with the time direction.

We have matched our deformed symmetry algebra with known results for the mirror model, \emph{despite} the subtleties surrounding fermions in the $q$-deformed $\ads$ model. The mirror model is a solution of supergravity, which at the bosonic level has $\kappa$-Poincar\'e type symmetry, and whose light cone supersymmetry algebra matches expectations from super $\kappa$-Poincar\'e type symmetry. Consistency of the full symmetry algebra strongly suggests that the full symmetry algebra of the mirror model has to be our tentative $\mathcal{U}_\kappa(\mathfrak{iusp}(2,2|4))$. As such, perhaps there is a version of the $q$-deformed $\ads$ model whose fermions contract directly to the mirror model, and at least in this limit would correspond to a solution of supergravity. Moving on, let us comment further on our contracted model, and related other ones.

\subsection{T duality}

Upon formally T dualizing in $t$ and $\phi$, our mirror $\ads$ becomes $\mathrm{dS}_5 \times \mathrm{H}^5$, the product of five-dimensional de Sitter space and a five-dimensional hyperboloid \cite{Arutyunov:2014cra}.\footnote{This fact was previously observed by S. Frolov.} In fact, if we keep the dependence on $\kappa$ as in eqn. \eqref{eq:mirrorwithkappa}, it becomes the radius of $\mathrm{dS}_5$ and $\mathrm{H}^5$.

Forgetting about fermions for a moment, we see that timelike T duality relates the sigma model on de Sitter space to the Lorentzian submanifold of mirror $\ads$, and at the level of symmetry thus apparently relates  $q$-deformed Poincar\'e symmetry to undeformed Lorentz symmetry in one dimension higher. Similarly, T duality in $\phi$ relates $\mathrm{H}^5$ to the Euclidean submanifold, and $q$-deformed Euclidean symmetry to undeformed Lorentz symmetry in one dimension higher. In fact, the $\mathrm{dS}_n$ and $\mathrm{H}^n$ sigma models should have Yangian symmetry, and we expect our $q$-deformed symmetry to extend to a full quantum affine algebra.

Including fermions we need both T dualities to get a clean statement. Taking us a bit beyond conventional strings, $\mathrm{dS}_5 \times \mathrm{H}^5$ is a solution of type IIB$^*$ supergravity \cite{Hull:1998vg}, and the corresponding superalgebra is a different real form of $\mathfrak{sl}(4|4)$ known as $\mathfrak{su}^*(4|4)$ \cite{Lukierski:1984it}. This double T duality hence appears to relate $\mathfrak{su}^*(4|4)$ and our tentative $\mathcal{U}_\kappa(\mathfrak{iusp}(2,2|4))$.

\subsection{Other dimensions, other spaces}

The $q$ deformation of \cite{Delduc:2013fga,Delduc:2013qra} applies to any $G/H$ (semi)symmetric space sigma model, and many of them are amenable to Wigner-\.{I}n\"on\"u contraction. Firstly however, we should come back to the option of using different $\mathrm{R}$ operators to deform $\ads$. By permuting the signature of $\mathfrak{su}(2,2)$, two other and apparently inequivalent deformations of $\mathrm{AdS}_5$ were constructed in \cite{Delduc:2014kha}. We briefly discuss these in appendix \ref{app:altdefAdSxS}. As explained there, they are not (directly) amenable to the contraction procedure we followed above.

Next, attempting to deform $\mathrm{dS}_5 \times \mathrm{H}^5$ analogously to $\ads$ as in the main text - which would straightforwardly contract to the T dual of $\mathrm{AdS}_5 \times \mathrm{S}^5$ - appears to conflict with the real form $\mathfrak{su}^*(4|4)$. It is possible to deform $\mathrm{dS}_5 \times \mathrm{H}^5$, but this results in the analog of one of the other deformations of $\ads$ just mentioned, see appendix \ref{app:deformeddSxH} for details.

Of course we can consider (anti-)de Sitter space, the sphere, or the hyperboloid in any dimension, with metrics corresponding to the obvious analogue of (parts of) eqn. \eqref{eq:mirrorwithkappa}. In particular, timelike four-dimensional $\kappa$-Poincar\'e symmetry arises in the four-dimensional sigma model obtained by contracting the deformed four-dimensional $\mathrm{AdS}_4$ sigma model, namely\footnote{J. Lukierski informed us that eqn. (19) as well as its counterpart for the spacelike deformation alluded to below have been independently obtained by A. Borowiec, H. Kyono, J. Lukierski, J. Sakamoto, and K. Yoshida \cite{Borowiec:2015wua}.}
\begin{align}
ds^2 = \,&
\frac{-dt^2  + d r^2}{1-r^2/\kappa^{2}} + r^2 d\Theta_{2}.\label{eq:4dmirror}
\end{align}
Analogously, we expect the T dual of four-dimensional anti-de Sitter space to come out of contracting the split type deformation of $\mathrm{AdS}_4$, which should realize spacelike $\kappa$-Poincar\'e symmetry, though we have only concretely investigated this in two dimensions, see appendix \ref{app:2dcases} for details. Similarly, split deformations exists for $\mathrm{AdS}_3$ and $\mathrm{AdS}_5$, which we expect to contract analogously. These cannot be lifted to the corresponding superstring coset sigma models however, since the associated spheres do not admit split deformations.

At least in two dimensions, we can also find sigma models with null type $\kappa$-Poincar\'e symmetry. As discussed in appendix \ref{app:2dcases} this uses a deformation of $\mathrm{AdS}_2$ given in \cite{vanTongeren:2015soa}. This deformation is based on the (homogeneous) classical Yang-Baxter equation (CYBE) instead of the mCYBE - which is also an option in this context \cite{Kawaguchi:2014qwa} - matching the fact that precisely in the null case the $\kappa$-Poincar\'e $r$ matrix satisfies the CYBE instead of the mCYBE (see e.g. \cite{Borowiec:2013lca}). This means it is associated with a Drinfeld twist and not a proper $q$ deformation, but in this case the contraction appears to work nicely as well. Let us also note that since $\mathfrak{so}(2,4)$ contains an $\mathfrak{iso}(1,3)$ subalgebra, it is possible to directly deform the $\ads$ string by the \emph{four}-dimensional null $\kappa$-Poincar\'e $r$ matrix, as considered in \cite{vanTongeren:2015uha}.\footnote{Generically CYBE based deformations of the $\ads$ string are conjectured to give gravity duals to various noncommutative versions of supersymmetric Yang-Mills theory \cite{vanTongeren:2015uha}.}

Finally, at the bosonic level it might seem nice to try to deform the flat space string to get a sigma model with ten-dimensional $\kappa$-Poincar\'e symmetry. This would result in a ten-dimensional analogue of the above metric. However, since it contains only one of the factors of mirror $\ads$, or equivalently since it is T dual to $\mathrm{dS}_{10}$, it cannot be embedded in standard supergravity.

\subsection{Deforming flat space directly}

We $q$ deformed the flat space string by contracting the $q$-deformed $\ads$ string, noting that semisimple groups - as opposed to the super Poincar\'e group of the flat space string - can be naturally $q$ deformed. However, as the $r$ matrix contracts nicely, and the flat space string can be viewed as a coset model on $(\mathcal{N}=2 \,\, \mathrm{ISO}(1,9))/\mathrm{SO}(1,9)$ \cite{Henneaux:1984mh}, it might be possible to directly deform the flat space string using this $\kappa$-Poincar\'e type $r$ matrix. Of course, in order to include the flat space fermions, we would first have to work out the supersymmetrized $\kappa$-Poincar\'e type $r$ matrix we have implicitly described here \emph{via a contraction}, in full detail. Still, at the bosonic level these $r$ matrices were already known, and we could attempt to proceed directly. Indeed, the deformed four-dimensional Minkowski space of eqn. \eqref{eq:4dmirror} was concurrently obtained in \cite{Borowiec:2015wua} in this fashion. In this approach we cannot escape the distinguished nature of nonsemisimple algebras however: their Killing form is degenerate. Because of this the Yang-Baxter coset sigma model approach does not immediately apply. In particular we need to construct an $\mathrm{R}$ operator from the $r$ matrix, requiring a nondegenerate symmetric invariant bilinear form. As advocated in \cite{Matsumoto:2015ypa,Borowiec:2015wua}, we can go around this by working with a larger semisimple algebra that contains the Poincar\'e algebra, such as the conformal algebra. One can then embed Minkowski space in the $\mathrm{AdS}_5$ coset model, and heuristically define the $\mathrm{R}$ operator, vielbeins, associated metric, and $B$ field via traces involving algebra elements not in the Poincar\'e algebra. The result agrees with ours above, however the approach remains heuristic, and it is not clear to what extent one should allow the enlarged algebra to be used. Allowing an $r$ matrix to involve the dilatation generator can apparently for instance result in exactly the same deformed model \cite{Matsumoto:2015ypa}, though the underlying algebraic interpretation should be very different because the associated $r$ matrix solves the CYBE rather than the mCYBE, and is thereby associated with a twisted conformal rather than $\kappa$-Poincar\'e type algebra. Our approach based on contraction allows us to avoid these subtleties. Furthermore, it is not obvious that one can always find a suitably larger coset model that one can ``embed'' the desired model with nonsemisimple symmetry in. In particular, if possible, for the flat space superstring this would require moving beyond ten dimensions, careful treatment of the fermions, and an $r$ matrix currently defined through contraction in any case.

\section{Outlook}

In this paper we discussed a contraction limit of the $q$ deformation of the $\ads$ superstring sigma model, which can be viewed as a $q$ deformation of the flat space string. Interestingly, this contraction limit turns out to be identical to the one used to obtain the so-called mirror background, showing that the light cone $\ads$ string is the double Wick rotation of the light cone gauge fixed $q$-deformed flat space string. Similar stories apply directly to $\mathrm{AdS}_2 \times \mathrm{S}^2 \times \mathrm{T}^6$ and $\mathrm{AdS}_3 \times \mathrm{S}^3 \times \mathrm{T}^4$, the only difference being that there the deformation involves smaller algebras. In this sense, the ``most $q$-deformed'' flat space superstring is obtained starting from ($q$-deformed) $\ads$, as it has the biggest semisimple superalgebra underlying a particular string. The backgrounds coming out of our contraction procedure are T dual to other (semi)symmetric spaces, indicating interesting relations between (infinite-dimensional) symmetry algebras of sigma models under T duality.

There are a number of interesting open questions that we did not address in this paper. Firstly, we did not explicitly contract the deformed algebra, or include the fermions beyond matching the central extension. We hope to address these points in the near future, in particular the relevant super $\kappa$-Poincar\'e algebra. On a related note, though involved, it would be very interesting to clarify the relationship between the possible versions of the $q$-deformed $\ads$ model and the mirror model at the fermionic level, as the latter is conformal at least at one loop. Beyond this, it might be interesting to concretely investigate (contractions of) split deformations of $\mathrm{AdS}_3$, $\mathrm{AdS}_4$ and $\mathrm{AdS}_5$, and their relation to spacelike $\kappa$-Poincar\'e symmetry. From the perspective of string theory, looking into an $r$ matrix that contracts to the five-dimensional null case is perhaps more interesting however. If this exists (it does in two dimensions) it should correspond to a Drinfeld twist, and can be applied to the full string. At the algebraic level this would give a twisted rather than $q$-deformed algebra, with a nontrivial contraction. It would also be interesting to further investigate the deformations of $\ads$ corresponding to the other choices of Drinfeld-Jimbo $r$ matrices given in \cite{Delduc:2014kha}, regardless of their (lack of) contractibility. Furthermore, in general it would be great to get a full grasp on the infinite-dimensional symmetry algebras of these models, even already at the bosonic level, and study their relation under T duality in detail. Moreover, it might be interesting to consider one-sided contractions in the two parameter deformation of $\mathrm{AdS}_3 \times \mathrm{S}^3 \times \mathrm{T}^4$ \cite{Hoare:2014oua}. Finally, it may be interesting to consider the Penrose limit of $\ads$ in this deformed setting.

\begin{acknowledgments}
We would like to thank B. Hoare, V. Giangreco M. Puletti, and A. Tseytlin for the insightful discussions, and G. Arutyunov, B. Hoare, and J. Lukierski for comments on the draft.  ST is supported by LT. The work of ST is supported by the Einstein Foundation Berlin in the framework of the research project "Gravitation and High Energy Physics" and acknowledges further support from the People Programme (Marie Curie Actions) of the European Union's Seventh Framework Programme FP7/2007-2013/ under REA Grant Agreement No 317089. AP acknowledges funding from the European Union's Seventh Framework Programme for research and innovation under the Marie Sk{\l }odowska-Curie grant agreement No 609402 - 2020 researchers: Train to Move (T2M). Part of this work is supported by Narodowe Centrum Nauki (Project No.:2014/13/B/ST2/04043).
\end{acknowledgments}

\appendix

\section{Algebra and canonical $r$ matrix}

\label{app:algebraandrmatrix}

In this paper we are mainly concerned with the bosonic subalgebra $\mathfrak{%
su}(2,2)\oplus\mathfrak{su}(4)$ of $\mathfrak{\mathfrak{psu}}(2,2|4)$. For
details and the supersymmetric extension of the material here, we refer to
the pedagogical review \cite{Arutyunov:2009ga} whose conventions we follow.
We will only briefly list the facts we need, beginning with the $\gamma$
matrices
\begin{equation}
\begin{aligned} &\gamma^0 = i \sigma_3 \otimes \sigma_0, &\gamma^1 =
\sigma_2 \otimes \sigma_2, &&\gamma^2 = -\sigma_2 \otimes \sigma_1, \\
&\gamma^3 = \sigma_1 \otimes \sigma_0, &\gamma^4 = \sigma_2 \otimes
\sigma_3, &&\gamma^5 = -i \gamma^0, \end{aligned}
\end{equation}
where $\sigma_0 = 1_{2\times2}$ and the remaining $\sigma_i$ are the Pauli
matrices. With these matrices the generators of $\mathfrak{so}(1,4)$ in the
spinor representation are given by $m^{ij} = \frac{1}{4} [\gamma^i,\gamma^j]$
where the indices run from zero to four, while for $\mathfrak{so}(5)$ we can
give the same construction with indices running from one to five. The
algebra $\mathfrak{su}(2,2)$ is spanned by these generators of $\mathfrak{so}%
(1,4)$ together with the $m^{i5}=\frac{1}{2}\gamma ^{i}$ for $i=0,\ldots,4$,
satisfying:%
\begin{equation}
\left[ m^{ij},m^{kl}\right] =\eta ^{jk}m^{il}-\eta ^{ik}m^{jl}-\eta
^{jl}m^{ik}+\eta ^{il}m^{jk}
\end{equation}
where $i,j,k,l=0,...,5$ and $\eta =\mathrm{diag}\left( -1,1,1,1,1,-1\right) $%
. While $\mathfrak{su}(4)$ is spanned by the combination of $\mathfrak{so}(5)
$ and $i\gamma^j$ for $j=1,\ldots,5$ with
\begin{equation}
\left[ n^{ij},n^{kl}\right] =\delta ^{jk}n^{il}-\delta ^{ik}n^{jl}-\delta
^{jl}n^{ik}+\delta ^{il}n^{jk}
\end{equation}
where $i,j=1,...,5$ and $n^{i6}=\frac{i}{2}\gamma ^{i}$. Concretely, these
generators satisfy
\begin{equation}  \label{realm}
m^\dagger \gamma^5 + \gamma^5 m = 0
\end{equation}
for $m\in \mathfrak{su}(2,2)$, and
\begin{equation}  \label{realn}
n^\dagger+ n = 0
\end{equation}
for $n \in \mathfrak{su}(4)$. This means that we are dealing with the
canonical group metric $\gamma^5 = \mathrm{diag}(1,1,-1,-1)$ for $\mathrm{SU}%
(2,2)$, and that $e^{\alpha n}$ and $e^{\alpha m}$ give group elements for
real $\alpha$.

The $\mathbb{Z}_4$ automorphism of $\mathfrak{\mathfrak{psu}}(2,2|4)$ is
generated by $\Omega$, which acts on the bosonic subalgebras as
\begin{equation}  \label{eq:Z4gendef}
\Omega(m) = -K m^t K,
\end{equation}
where $K=-\gamma^2\gamma^4$. It leaves the above mentioned subalgebras $%
\mathfrak{so}(1,4)$ and $\mathfrak{so}(5)$ invariant.

The deformation of the above algebras is normally introduced in the
Cartan-Chevalley basis for the complexified algebras, so let us give this as
well. Here $h_{i}$ are Cartan elements and for simple roots $\alpha _{i}$ we
have the corresponding single vectors $e_{i}=e_{+\alpha
_{i}},f_{i}=e_{-\alpha _{i}}$. They satisfy
\begin{eqnarray}
\left[ h_{i},h_{j}\right]  &=&0  \label{h} \\
\left[ e_{i},f_{j}\right]  &=&\delta _{ij}h_{i}  \label{ef} \\
\left[ h_{i},e_{j}\right]  &=&a_{ij}e_{j}\quad ,\quad \left[ h_{i},f_{j}%
\right] =-a_{ij}f_{j}  \label{he}
\end{eqnarray}%
where $a_{ij}$ is the Cartan matrix. After deformation\footnote{%
To be precise the deformation procedure requires working with the universal
enveloping algebra $\mathcal{U}(\mathfrak{g})$ of a given Lie algebra $%
\mathfrak{g}$ and besides relations (\ref{h}-\ref{he}) one needs to take
into account the so-called Serre relations which also become $q$-deformed.
There is a unique Hopf algebra structure on the $q$-deformed algebra $%
\mathcal{U}(\mathfrak{g})$, see e.g. Sec. 6 in \cite{Klimyk:1997eb}.} the
commutation relation (\ref{ef}) becomes
\begin{equation}
\left[ e_{i},f_{j}\right] =\delta _{ij}[h_{i}]_{q},
\end{equation}%
and the relations for the non-simple root vectors (\ref{nonsimp}) become $q$%
-commutators. For the algebras $\mathfrak{su}(2,2)$ (and $\mathfrak{su}(4)$)
we take the Cartan subalgebra to be spanned by
\begin{eqnarray}
h_{1} &=&\mathrm{diag}(1,-1,0,0),\quad \quad h_{2}=\mathrm{diag}(0,1,-1,0),
\notag \\
h_{3} &=&\mathrm{diag}(0,0,1,-1).
\end{eqnarray}%
The Cartan matrix is
\begin{equation}
a_{ij}=\left(
\begin{array}{ccc}
2 & -1 & 0 \\
-1 & 2 & -1 \\
0 & -1 & 2%
\end{array}%
\right) .
\end{equation}%
The non-simple root vectors follow from
\begin{align}
e_{4} &=i[e_{1},e_{2}],& f_{4}&=i[f_{1},f_{2}],  \notag
\label{nonsimp} \\
e_{5} &=i[e_{2},e_{3}],& f_{5}&=i[f_{2},f_{3}], \\
e_{6} &=i[e_{1},i[e_{2},e_{3}]],& f_{6}&=i[f_{1},i[f_{2},f_{3}]].
\notag
\end{align}%
The extended Cartan elements are
\begin{equation}
h_{4}=h_{1}+h_{2},\quad h_{5}=h_{2}+h_{3},\quad h_{6}=h_{1}+h_{2}+h_{3}.
\notag
\end{equation}%
We have raising generators $e_{i}$ and lowering generators $f_{i}$, $%
i=1,\ldots ,6$. The relations between the (bosonic) generators of $\mathfrak{%
su}(2,2)$ (and $\mathfrak{su}(4)$) and the Cartan basis elements are
{\footnotesize {\
\begin{eqnarray}
m^{05} &=&\frac{i}{2}\left( h_{1}+2h_{2}+h_{3}\right) =n^{56},  \notag \\
m^{12} &=&\frac{i}{2}\left( h_{1}+h_{3}\right) =n^{12},  \notag \\
m^{34} &=&\frac{i}{2}\left( h_{1}-h_{3}\right) =n^{34},  \notag \\
n^{13} &=&\frac{i}{2}\left( e_{1}+f_{1}-e_{3}-f_{3}\right) =m^{13},  \notag
\\
n^{14} &=&\frac{1}{2}(e_{1}-f_{1}+e_{3}-f_{3})=m^{14},  \notag \\
n^{23} &=&\frac{1}{2}(e_{1}-f_{1}-e_{3}+f_{3})=m^{23},  \notag \\
n^{24} &=&-\frac{i}{2}(e_{1}+f_{1}+e_{3}+f_{3})=m^{24},  \notag \\
n^{15} &=&\frac{i}{2}(e_{2}+f_{2}-e_{6}-f_{6})=im^{01},  \notag \\
n^{25} &=&\frac{1}{2}(-e_{2}+f_{2}-e_{6}+f_{6})=im^{02},  \notag \\
n^{16} &=&\frac{1}{2}(e_{2}-f_{2}-e_{6}+f_{6})=im^{15},  \notag \\
n^{26} &=&\frac{i}{2}\left( e_{2}+f_{2}+e_{6}+f_{6}\right) =im^{25},  \notag
\\
n^{35} &=&-\frac{i}{2}\left( e_{4}+e_{5}+f_{4}+f_{5}\right) =im^{03},  \notag
\\
n^{45} &=&\frac{i}{2}\left( -e_{4}+e_{5}+f_{4}-f_{5}\right) =im^{04},  \notag
\\
n^{36} &=&\frac{1}{2}\left( -e_{4}-e_{5}+f_{4}+f_{5}\right) =im^{35},  \notag
\\
n^{46} &=&-\frac{i}{2}\left( -e_{4}+e_{5}-f_{4}+f_{5}\right) =im^{45}.
\notag
\end{eqnarray}%
} }
\normalsize  Note that we have $h_{i}^{\dagger }=h_{i}$ and
$e_{i}^{\dagger }=f_{i}$.

The standard Drinfeld-Jimbo classical $r$ matrix $r\in\wedge
^{2}\mathfrak{g}$ for $\mathfrak{so}\left( 2,4\right) \sim\mathfrak{su}%
\left( 2,2\right) $ is
\begin{align}
r_{DJ}&=i \sum_{j=1}^6 e_j \wedge f^j  \label{eq:rDJ} \\
&=(m_{0i}\wedge m^{i5}+m_{23}\wedge m^{13}+m_{24}\wedge m^{14}),  \notag
\end{align}
sums in $i$ running from zero to five. Note that here and below the $i=0$
and $i=5$ terms give zero however. This $r$ matrix satisfies the
inhomogeneous (modified) Yang-Baxter equation
\begin{equation}  \label{eq:rmatmCYBE}
[r_{12},r_{13}]+[r_{12},r_{23}]+[r_{13},r_{23}]=\Omega
\end{equation}
with the invariant $\Omega =m_{j}^{i}\wedge m_{k}^{j}\wedge m_{i}^{k} \in
\wedge ^{3}\mathfrak{so}\left( 2,4\right)$. The corresponding $\mathrm{R}$
operator is defined by
\begin{equation}  \label{eq:rDJOp}
\mathrm{R}(x) \equiv \mbox{sTr}_2(r_{DJ} (1\otimes x)),
\end{equation}
where now eqn. \eqref{eq:rmatmCYBE} is equivalent to the operator form of
the mCYBE of eqn. \eqref{eq:ROpmCYBE} via the nondegenerate Killing form of $\mathfrak{so}(2,4)$.

Redefining $m_{i5}=Rp_i$, eqn. (\ref{eq:rDJ}) becomes
\begin{equation}
r_{DJ}= R(m_{0i}\wedge p^{i}+\frac{1}{R}m_{23}\wedge m^{13}+\frac{1}{R}%
m_{24}\wedge m^{14})
\end{equation}
At this stage it is convenient to factor in the overall deformation
parameter $\varkappa = R \kappa^{-1}$, so that $\varkappa \, r_{DJ}$
contracts to
\begin{equation}  \label{contr-m}
\kappa^{-1} r_\kappa:=\lim_{R\rightarrow \infty}\frac{\varkappa}{R^2} \,
r_{DJ}(\frac{1}{R})=\kappa^{-1} m_{0i}\wedge p^{i}.
\end{equation}
For $\mathfrak{su}(4)$ we analogously have
\begin{equation}
r_{DJ}=(n_{5i}\wedge n^{i6}+n_{23}\wedge n^{31}+n_{14}\wedge n^{42}),
\end{equation}
which contracts to
\begin{equation}  \label{contr-n}
r_\kappa= n_{5i}\wedge l^{i}.
\end{equation}
The $\mathrm{R}$ operator referred to in the main text corresponds to the
standard $\mathfrak{psu}(2,2|4)$ $r$ matrix, which at the bosonic level is
just the combination of the above $\mathfrak{so}(2,4)$ and $\mathfrak{so}(6)$
$r$ matrices.

\section{Alternate $q$ deformations of $\ads$}
\label{app:altdefAdSxS}

The deformation of $\ads$ described in the main text is not the only possibility within the framework of \cite{Delduc:2013qra,Delduc:2014kha}. In \cite{Delduc:2014kha} it was shown that by combining the $\mathrm{R}$ operator with permutations, it is possible to obtain two other deformations of $\mathrm{AdS}_5$ \cite{Delduc:2014kha}. This procedure essentially amounts to permuting the signature of $\mathfrak{su}(2,2)$, which does nothing for $\mathfrak{su}(4)$ and hence should not give further deformations of the sphere. We believe that these three inequivalent permutations correspond to the three possible choices for Cartan involutions for the real form $\mathcal{U}_q(\mathfrak{su}(2,2))$ with real $q$ described in \cite{Lukierski:1992ex}, cf. eqs. (3.5) and table 1 there.

In principle it is possible to contract these different forms of $\mathcal{U}_q(\mathfrak{so}(2,4))$. In terms of physical generators, the two $r$ matrices corresponding to permutation $\mathbb{P}_1$ and $\mathbb{P}_2$ in \cite{Delduc:2014kha} are given by
\begin{equation}
r_1 =  m_{1i} \wedge m^{i2} - m_{03} \wedge m^{35} - m_{04} \wedge m^{45},
\end{equation}
and
\begin{equation}
r_2 =  m_{3i} \wedge m^{i4} + m_{10} \wedge m^{02} + m_{15} \wedge m^{52}
\end{equation}
respectively. At this stage an important difference to $r$ of eqn. \eqref{eq:rDJ} becomes clear: the obvious analogous contraction should now involve directions one or two for $r_1$, or three and four for $r_2$, however these are in the direction of the denominator of the coset $\mathrm{SO}(2,4)/\mathrm{SO}(1,4)$ (gauge directions). It is therefore not clear to us that the corresponding contraction limit can be simply implemented in the sigma model.

\section{Deformed $\mathrm{dS}_5 \times \mathrm{H}^5$}
\label{app:deformeddSxH}

In this appendix we briefly describe how to find a deformation of $\mathrm{dS}_5 \times \mathrm{H}^5=\mathrm{SO}(1,5)/\mathrm{SO}(1,4) \times\mathrm{SO}(1,5)/\mathrm{SO}(5) $ in the spirit of \cite{Delduc:2013qra,Arutyunov:2013ega}. It will however not be the analogue of the deformation of $\ads$ in the main text, rather it is the analogue of the deformation obtained from the $\mathrm{R}$ operator associated with $r_2$ above. The reason for this is that the standard Drinfeld-Jimbo $r$ matrix of $\mathfrak{sl}(4|4)$ (with respect to our basis) does not preserve the real form $\mathfrak{su}^*(4|4)$.\footnote{We can of course forget about reality and ask what happens if we use this canonical $r$ matrix to deform $\mathrm{dS}_5 \times \mathrm{H}^5$. Naturally this gives the analytic continuation  $t \rightarrow i t, \rho \rightarrow i \rho$, and $\phi \rightarrow i \phi$ and $r \rightarrow i r $ of deformed $\ads$ described above. This appears to be a real model. However, formally the construction also yields an imaginary exact term in the $B$ field, cf. the continuation of the one $\ads$, see e.g. footnote 17 in \cite{Arutyunov:2015qva}. It would be nice to understand this point in full detail.} Without loss of generality, let us describe the situation at the level of $\mathfrak{sl}(4)$ and $\mathfrak{su}^*(4)$. Following \cite{Delduc:2014kha} we consider permutations of $\{1,2,3,4\}$, and use them to construct possibly inequivalent $\mathrm{R}$ operators. In our case there turns out to be only one inequivalent permutation that yields an $\mathrm{R}$ operator compatible with $\mathfrak{su}^*(4)$.\footnote{Note that this counting appears to match with the results of \cite{Lukierski:1992ex}, cf. eqs. (3.5) and table 1 there, with three different real $q$ deformations of $\mathfrak{so}(2,4)$, but only one for $\mathfrak{so}(1,5)$.} We can take this permutation to be
\begin{equation}
\mathbb{P} = \left(\begin{array}{cccc}1&2&3&4\\1&4&2&3 \end{array}\right)
\end{equation}
where the corresponding $\mathrm{R}$ operator $\mathrm{R}^\mathcal{P}$ is given by $\mathrm{Ad}_{\mathcal{P}^{-1}} \circ \mathrm{R} \circ \mathrm{Ad}_\mathcal{P}$ with $\mathcal{P}_{ij}=\delta_{i\mathbb{P}(j)}$, while $\mathrm{R}$ denotes the canonical Drinfeld-Jimbo $\mathrm{R}$ operator of eqn. \eqref{eq:rDJOp}.

The metric and $B$ field corresponding to $\mathrm{R}^\mathcal{P}$ are given by taking those of the deformed $\ads$ corresponding to $\mathbb{P}^2$ of appendix D.1 of \cite{Delduc:2014kha} and analytically continuing $t \rightarrow i t, \rho \rightarrow i r$, and $\phi \rightarrow i \phi$ and $r \rightarrow i \rho $. It can be constructed directly by analytically continuing the generators as $m^{i5}\rightarrow \hat{m}^{i5} = i m^{i5}$ and $n^{i6}\rightarrow \hat{n}^{i6} = -i n^{i6}$ (sign choices of course do not affect the outcome), and using group elements of the form
\begin{align*}
 g_d & = e^{\psi^i \hat{f}_i} e^{\zeta \hat{m}^{13}} e^{\arcsin{\rho}\, \hat{m}^{15}},\\
 g_h & = e^{\chi^i \hat{f}_i} e^{\xi \hat{n}^{13}} e^{\arcsinh{r}\, \hat{n}^{16}},
\end{align*}
for de Sitter space ($d$) and the hyperboloid ($h$) respectively, where the $\hat{f}$ generators are related to the $\hat{m}$ and $\hat{n}$ generators as the $f$ are to $m$ and $n$. The algebra $\mathfrak{su}^*(4)$ as obtained this way is spanned by the matrices $y$ that satisfy $\hat{K}^{-1}y^*\hat{K} =y$, with $\hat{K}=\mbox{diag}(i\sigma_2,-i\sigma_2)$ (matching the conventions of \cite{Delduc:2014kha}).

Like the alternate $r$ matrix deformations of $\ads$ discussed in the appendix above, the two obvious sensible contractions of the $r$ matrix for this deformed model would use coset denominator (gauge) directions at the level of the sigma model.

\section{Deforming $\mathrm{S}^2$, $\mathrm{AdS}_2$, $\mathrm{dS}_2$, and $\mathrm{H}^2$}

\label{app:2dcases}

The restrictions imposed by the larger algebras involved for $\ads$ and $\mathrm{dS}_5 \times \mathrm{H}^5$ are not necessarily present in lower-dimensional algebras. Let us therefore briefly consider the two-dimensional sphere and its various analytic continuations. When analytically continuing it becomes possible to have split $r$ matrices. Let us begin the discussion with the sphere, i.e. $\mathfrak{su}(2)$ at the algebraic level.

With respect to our conventions, the standard nonsplit $\mathrm{R}$ operator in this case is associated with the $r$ matrix $r^{(12)}= T_1 \wedge T_2$, where the $T_i, i=1,2,3$, denote the generators of $\mathfrak{su}(2)$. The coset denominator $U(1)$ is associated with $T_2$, while in our conventions the coordinates $r$ and $\phi$ of the main text are associated with $T_1$ and $T_3$ respectively. Of course, these preferred directions have no meaning in the algebra, and any group rotation of the $r$ matrix is still a solution of the mCYBE. For our purposes it will be useful to consider the permutations $r^{(13)}=T_1 \wedge T_3$ and $r^{(23)}=T_2 \wedge T_3$. The associated deformations of $\mathrm{S}^2$ are equivalent, but do not manifestly appear so in terms of coordinates that each have relevance after analytic continuation. We have
\begin{align}
ds^2_{12} &  = \frac{1}{1+\varkappa^2 r^2}d\Omega_2^2 ,\\
ds^2_{13} &  = \frac{2}{2+\varkappa^2(1+\cos 2 \phi - \cos^2 \phi\, r^2)}d\Omega_2^2 ,\\
ds^2_{23} &  = \frac{2}{2+\varkappa^2(1-\cos 2 \phi - \sin^2 \phi\, r^2)}d\Omega_2^2 ,
\end{align}
all up to total derivative B fields. Here $d\Omega_2^2$ denotes the metric on the two sphere in $r$ and $\phi$ coordinates analogous to the ones in the main text. Clearly these last two are related by the shifting $\phi$ by $\pi/2$, the relation between the first and the two others is not as clear.

From here we can directly derive the corresponding deformations of $\mathrm{AdS}_2$, $\mathrm{dS}_2$ and $\mathrm{H}^2$ by appropriately analytically continuing $r$ and $\phi$ as well as the deformation parameter, including a factor of $i$ when the corresponding generator becomes noncompact, or the $r$ matrix changes from the nonsplit to the split type, cf. table \ref{tab:2doverview}. This table gives the possibilities that are compatible with the real form under consideration, meaning $r$ matrices containing only real combinations of generators.  Since formally multiplying $r$ by $i$ takes a solution of the nonsplit type, to one of the split type, it is easy to determine whether $r^{(ij)}$ is of split or nonsplit type, by comparing to the original nonsplit $\mathfrak{su}(2)$ $r$ matrix - if both $i$ and $j$ are compact or noncompact, it is of the nonsplit type, otherwise it is of the split type.\footnote{The mention in footnote \ref{footnote:Uqsl2} that in order to $q$ deform $\mathfrak{sl}(2)$ in a `standard' fashion, we need to consider $|q|=1$, while for $\mathfrak{su}(1,1)$ we would like $q$ real, seems to match the change from split to nonsplit $r$ matrix here.} Due to the analytic continuations involved, the resulting deformed geometries are not all equivalent under real diffeomorphisms, and may moreover involve analytically continuing $\varkappa$. If we would like to embed these algebras in a superalgebra, it is of course necessary to require that both $r$ matrices are of the same type, though it need not be sufficient.

\begin{table}
\begin{center}
\begin{tabular}{| c | c | c | c | c | c | c |}
  \hline
 & $T_1$ & $T_2$ & $T_3$ & $r^{(12)}$ & $r^{(13)}$ & $r^{(23)}$\\
 \hline
$\mathrm{S}^2$ & c & c & c & ns & ns & ns\\
\hline
$\mathrm{AdS}^2$ & nc & nc & c & ns & s & s \\
\hline
$\mathrm{dS}^2$ &  c  & nc & nc & s & s & ns\\
\hline
$\mathrm{H}^2$ &  nc  & c & nc & s & ns  & s\\
  \hline
  coord & $r,\rho$ & gauge & $\phi,t$ & & & \\
  \hline
\end{tabular}
\caption{Three possible $r$ matrices for each of $\mathrm{S}^2$, $\mathrm{AdS}^2$, $\mathrm{dS}^2$, and $\mathrm{H}^2$. The $T_i$ denote the three generators of the relevant real form of $\mathfrak{sl}(2,\mathbb{C})$, which can be associated with compact ($c$) or noncompact ($nc$) directions. The associated coordinate type is indicated at the bottom of the table. The $r$ matrices are of either split ($s$) or nonsplit ($ns$) type.}
\label{tab:2doverview}
\end{center}
\end{table}

Let us now consider possible interesting contractions of these models. As mentioned before, we want to contract in the non gauge directions, and the $r$ matrix should have a single component in there.\footnote{If we were to take two components in there, we would formally contract the $r$ matrix to zero under the same scaling. It appears that the associated sigma models simply vanish in this strict limit, and factoring out an overall vanishing scale the leading term is the associated flat space model, though possibly with a flipped signature.} This leaves us with $r^{(12)}$ and $r^{(23)}$. The result of the contractions is collected in table \ref{tab:2dcontractions}. Given the various relations under analytic continuation, the fact that some of the contractions come out the same is not too surprising - the contraction can trivialize the distinguishing effect.

\begin{table}
\begin{center}
\begin{tabular}{| c | c | c |}
  \hline
 & $r^{(12)}$ & $r^{(23)}$\\
 \hline
$\mathrm{S}^2$ & $T(\mathrm{H}^2)$ & $T(\mathrm{H}^2)$  \\
\hline
$\mathrm{AdS}^2$ & $T(\mathrm{dS}_2)$  & $T(\mathrm{AdS}_2)$ \\
\hline
$\mathrm{dS}^2$ &  $T(\mathrm{dS}_2)$  & $T(\mathrm{AdS}_2)$ \\
\hline
$\mathrm{H}^2$ & $T(\mathrm{H}^2)$  & $T(\mathrm{H}^2)$ \\
  \hline
\end{tabular}
\caption{Nontrivial contractions of the various deformed spaces. All contractions come out as T dual to some space $M$ ($T(M)$). The T dual spaces in the first column are obtained in coordinates analogous to the ones in the main text, the T duality referring to the isometry coordinate. Those in the second column are the spaces obtained by analytically continuing both coordinates to imaginary values (which simply flips the overall signature), and then T dualizing. E.g. $T(\mathrm{AdS}_2)$ means $ds^2 = \frac{-dt^2 + d\rho^2}{1-t^2}$.}
\label{tab:2dcontractions}
\end{center}
\end{table}

Note that at the level of $r$ matrices, the contraction of the alternate deformation of $\mathrm{AdS}_2$ - giving its own T dual - gives precisely the spacelike $\kappa$-Poincar\'e type $r$ matrix for $\mathfrak{iso}(1,1)$, namely
\begin{equation}
r^{(23)}_{\mathrm{AdS}_2} \rightarrow m_{01} \wedge p^{0}.
\end{equation}
Given that $\mathfrak{so}(2,2)$, $\mathfrak{so}(2,3)$ and $\mathfrak{so}(2,4)$ admit split $r$ matrices \cite{Cahen:1994} (see also \cite{Vicedo:2015pna} for a discussion in the present context), it should be possible to find corresponding split deformations of $\mathrm{AdS}_3$, $\mathrm{AdS}_4$, and $\mathrm{AdS}_5$, which we expect to contract to the T duals of themselves. These T dual models should then realize spacelike $\kappa$-Poincar\'e symmetry.

Finally, we should address the null $\kappa$-Poincar\'e case. In two dimensions this can be done by taking the deformation of $\mathrm{AdS}_2$ given in the conclusions of \cite{vanTongeren:2015soa} (fixing $\theta=\pi/2$), and contracting. This results in
\begin{equation}
ds^2 = \frac{-dt^2 +d\rho^2}{1-t^2 - \rho^2 + 2 t \rho} = \frac{dx^+ dx^-}{1-(x^-)^2},
\end{equation}
where we introduced light cone coordinates $x^\pm = \rho \pm t$. The associated sigma model hence should have two-dimensional null $\kappa$-Poincar\'e symmetry. For concreteness, the associated $r$ matrix is $r=T_1 \wedge T_2 + T_3 \wedge T_2$ (a solution of the CYBE instead of the mCYBE) which contracts exactly to the null $\kappa$-Poincar\'e $r$ matrix in two dimensions. It would be interesting to understand if similar $r$ matrices can be found that contract to null $\kappa$-Poincar\'e $r$ matrices in three, four, or five dimensions.

\bibliographystyle{apsrev4-1}
\bibliography{Stijnsbibfile}

\end{document}